\documentclass[12pt]{article}
\usepackage{epsfig,amsfonts,amssymb}
\usepackage{cite}
\input epsf.sty
\usepackage{cite}

\def\ZZZ{{\hbox{ Z\kern-1.6mm Z}}}
\def\RRR{{\hbox{ R\kern-2.4mm R}}}
\def\CCC{{\hbox{ C\kern-2.0mm C}}}
\def\zzz{{\hbox{z\kern-1mm z}}}

\newcommand{\qeq}{{\hbox{=\kern-2.3mm ? \kern.5mm }}}
\renewcommand{\qeq}{=}

\newcommand{\ws}{{\wt\sigma}}
\newcommand{\wrh}{{\wt\rho}}
\newcommand{\wv}{{\wt v}}

\newcommand{\CC}{{\cal C}}

\newcommand{\OO}{{\cal O}}

\newcommand{\wt}{\widetilde}
\newcommand{\wh}{\widehat}

\newcommand{\NN}{{\cal N}}

\newcommand{\be}{\begin{equation}}
\newcommand{\ee}{\end{equation}}
\newcommand{\ben}{\begin{eqnarray}\displaystyle}
\newcommand{\een}{\end{eqnarray}}

\newcommand{\bea}[1]{\begin{eqnarray}\label{#1} }
\newcommand{\eea}{\end{eqnarray}}

\newcommand{\refb}[1]{(\ref{#1})}
\newcommand{\p}{\partial}
\newcommand{\sectiono}[1]{\section{#1}\setcounter{equation}{0}}

\def\one{{\hbox{ 1\kern-.8mm l}}}
\def\zero{{\hbox{ 0\kern-1.5mm 0}}}

\begin{document}

\begin{center}
{\Large \bf
Two Centered Black Holes and N=4 Dyon Spectrum}

\end{center}

\vskip .6cm
\medskip

\vspace*{4.0ex}

\centerline{\large \rm  }

\vspace*{4.0ex}

\centerline{\large \rm Ashoke Sen}

\vspace*{4.0ex}

\centerline{\large \it Harish-Chandra Research Institute}

\centerline{\large \it  Chhatnag Road, Jhusi,
Allahabad 211019, INDIA}

\vspace*{1.0ex}
\centerline{E-mail:  sen@mri.ernet.in}

\vspace*{5.0ex}

\centerline{\bf Abstract} \bigskip

The exact spectrum of dyons in a class of N=4 supersymmetric 
string theories is known to change discontinuously across 
walls of  marginal stability. We show that the change in the
degeneracy  across the walls of
marginal stability can be accounted for precisely by the entropy
of two centered small black holes which (dis)appear
as we cross the walls of marginal stability.

\vfill \eject

\baselineskip=18pt

\tableofcontents

\sectiono{Introduction} \label{s1} 

We now have a good understanding
of the exact spectrum of a class of
quarter BPS dyons in $\NN=4$
supersymmetric string 
theories,  
obtained
by taking an asymmetric 
$\ZZZ_N$ orbifold of heterotic or
type IIA string theory compactified
on $T^6$\cite{9607026,0505094,0506249,0510147,0602254,
0603066,0605210,0607155,0609109,0612011,0702141,
0702150,0705.1433}. 
It is also known that as we cross
various walls of marginal stability associated with the possible
decay of the dyon into a pair of half BPS states,
the degeneracy changes by a
certain amount that is exactly computable\cite{0702141}. On the
other hand the asymptotic expansion of the degeneracy formula
for large charges reproduces the entropy of the corresponding
black hole not only to the leading order, but also to the first
subleading order in an expansion in inverse power of the 
charges\cite{0412287,0510147,0605210,0607155,0609109}. Given
this correspondence between dyon spectrum and black hole entropy,
a natural question to ask would be: can we understand the
jump in the degeneracy across walls of marginal stability on the
black hole side?

The question is somewhat tricky since these 
jumps in the degeneracy are
exponentially small compared to the leading contribution
to the entropy\cite{0702141}. 
Nevertheless since the change is 
discontinuous, one might hope that there is a clear mechanism on the
black hole side which produces these discontinuous changes
across  the walls of marginal stability
and if we can
identify this mechanism then we may be able to reproduce these jumps
on the black hole side. In this paper we shall show that there is
indeed a clear mechanism on the black hole side that describes these
jumps, -- this is the phenomenon of (dis)appearance of multicentered
black hole solutions for a given total charge as we cross various
walls of marginal stability in the space of asymptotic values of the
moduli fields\cite{0005049,0101135,0304094,0702146,0705.2564}.
In particular the exponential of the entropy 
associated with these multi-centered black holes will reproduce
the jump in the degeneracy computed from the exact dyon spectrum.

The role of multi-centered black holes in the context of exact dyon
spectrum of $\NN=4$ supersymmetric string theories has been discussed
before in \cite{0702150}. In this paper the authors considered
a special class of dyonic states for
which there is no single centered black hole solution but whose
degeneracy is predicted to be non-zero by the exact formula, and
showed how such states may be represented as 2-centered black holes.
However
for the charge vector used in \cite{0702150} each of these two
black holes had entropy of order unity, and hence their role in
producing the correct contribution to 
the degeneracy was not manifest.
In contrast we consider a dyonic state
with large charges for which the change in the degeneracy across the
wall of marginal stability is exponentially
large (even though it is exponentially
small compared to the leading contribution). The
2-centered black hole whose (dis)appearance across the wall of marginal
stability is responsible for this jump is a pair of small black holes
each carrying large charges and hence large 
entropy\cite{0409148,0410076,0411255,0411272,0501014,0502126,
0502157,
0507014,0506176}. Thus one 
can calculate the entropy 
associated with
this two centered black hole by using standard techniques and
compare it with the logarithm of the jump in the degeneracy across
the walls of marginal stability. The result turns out to be 
a perfect agreement.

\sectiono{Prediction from Exact Dyon Spectrum} \label{s2}

Our starting point will be heterotic or type IIA string theory compactified
on $T^4\times \wh S^1\times S^1$ modded out by a $\ZZZ_N$ group.
The action of the $\ZZZ_N$ group involves $1/N$ unit of translation
along $S^1$, together with an order $N$ transformation acting on the
degrees of freedom associated with $T^4$ and also (in the case of heterotic
theory) on the internal left-moving degrees of freedom. The $\ZZZ_N$
action is chosen so that it commutes with all the supersymmetries 
appearing from the right-moving sector of the world-sheet but (in case
of type IIA string theory) projects
out all the supersymmetries coming from the left-moving sector. In
this theory we shall consider dyons carrying momentum 
$(n',\wh n)$, winding $(-w',-\wh w)$,
Kaluza-Klein monopole charges $(N',\wh N)$
and H-monopole charges $(-W',-\wh W')$ along $S^1$ and 
$\wh S^1$ respectively. 
Such a dyon will be 
labelled by the electric and
magnetic charge vectors
\be \label{e1}
Q=\pmatrix{\wh n\cr n'\cr \wh w\cr w'}, \qquad 
P = \pmatrix{\wh W\cr W'\cr
\wh N\cr N'}\, .
\ee
The precise sign convention used for defining these charges can be
found in \cite{0705.1433}. We shall denote by $M$ the symmetric
$SO(2,2)$ matrix that encodes information about the moduli
labelling the torus $\wh S^1\times S^1$ and by $a+iS$ the axion-dilaton
modulus. If we denote by the subscript $_\infty$ the asymptotic values
of the various moduli, then the quarter BPS
dyon of charge $(Q,P)$ can decay into a pair of
half BPS states of charges $(Q,0)$ and $(0,P)$ on the wall of marginal
stability\cite{0702141}:
\be \label{e2}
a_\infty + {  (P^T (M_\infty + L) Q) \over 
\left[ (Q^T (M_\infty + L) Q) (P^T (M_\infty + L) P)
- (P^T (M_\infty + L) Q)^2\right]^{1/2}
}\, S_\infty =0\, ,
\ee
where
\be \label{e3}
L = \pmatrix{ 0 & I_2\cr I_2 & 0}
\ee
is the $SO(2,2)$ invariant matrix. 
There are other walls of marginal stability associated with the decay
into other pairs of half-BPS states\cite{0702141} 
but we shall carry out our
analysis in the vicinity of the wall
\refb{e2}. Other cases may be analyzed in the same way.

We shall consider diagonal $M_\infty$ of the form:
\be \label{e4}
M_\infty = \pmatrix{\wh R^{-2} &&&\cr & R^{-2} &&\cr && 
\wh R^2 &\cr
&&& R^2}\, .
\ee
In this case $\wh R$ and $R$ can be interpreted as the
radii of $\wh S^1$ and $S^1$ respectively, measured in 
units of $\sqrt{\alpha'}$. We shall also focus
on a special class of dyons for which\footnote{By following the
procedure given in \cite{0705.1433} we could switch on non-zero
values of the first and third components of $Q$, but in order to
keep the various formul\ae\ simple we shall continue to work with
the charge vector given in \refb{e5}.}
\be \label{e5}
Q =\pmatrix{0\cr -n/N \cr 0 \cr -1}, \qquad P = \pmatrix{Q_1-1\cr
-J\cr Q_5 \cr 0}\, , \qquad n, J, Q_1, Q_5\in \ZZZ, \quad
n, Q_1\ge 0, \quad Q_5 > 0\, ,
\ee
since for these states the exact degeneracy -- more precisely an index
that counts the number of bosonic minus the number of fermionic
supermultiplets\footnote{The degeneracy $d(\vec Q, \vec P)$
given in \refb{e7} actually refers 
to
the number of bosonic minus fermionic supermultiplets multiplied
by a factor of $(-1)^{Q\cdot P+1}$. The $(-1)^{Q\cdot P+1}$ factor
was not included in the analysis of 
\cite{0605210,0607155,0609109,0702141}. The $(-1)^{Q\cdot P}$ factor 
appeared in \cite{0508174} and reflects the change in statistics in going 
from a five to four dimensional viewpoint in the presence of a 
Kaluza-Klein monopole. The additional $-$ sign appears in the study of the 
bound state of a D1-D5 system to a Kaluza-Klein 
monopole\cite{pope,9912082}. These will be discussed in detail in a
forthcoming review\cite{appear}. \label{f1}}
-- can be computed by using a dual type IIB
description\cite{0605210,0607155,0609109}.
In this case  \refb{e2} takes the form:
\be \label{e6}
a_\infty = a_c, \quad
a_c \equiv -{J \, \wh R\over R \{Q_1 - 1 + \wh R^2 Q_5\}}\,
S_\infty\, .
\ee
The weak coupling region of the dual type IIB string theory
corresponds to the large $R$ region in the current 
description\cite{0702141}. In
this region the 
degeneracy formula takes the 
form\cite{0605210,0607155,0609109} 
(see \cite{0702141} for a review
of the results):
\be \label{e7}
d(Q,P) = \cases {d_>(Q,P) \quad \hbox{for $a_\infty> a_c$}\cr 
d_<(Q,P) \quad \hbox{for $a_\infty< a_c$}}\, ,
\ee
where
\ben \label{e8}
d_>(\vec Q,\vec P) &=& {1\over N}\, \int _{\CC_>} 
d\wt\rho \, 
d\wt\sigma \,
d\wt v \, e^{-\pi i ( N\wt \rho Q^2
+ \wt \sigma P^2/N +2\wt v Q\cdot P)}\, {1
\over \wt\Phi(\wt \rho,\wt \sigma, \wt v)}\, , \nonumber \\
d_<(\vec Q,\vec P) &=& {1\over N}\, \int _{\CC_<} 
d\wt\rho \, 
d\wt\sigma \,
d\wt v \, e^{-\pi i ( N\wt \rho Q^2
+ \wt \sigma P^2/N +2\wt v Q\cdot P)}\, {1
\over \wt\Phi(\wt \rho,\wt \sigma, \wt v)}\, .
\een
Here 
\be \label{edefq2}
Q^2=Q^TLQ =2n/N, \qquad
P^2=P^TLP=2Q_5(Q_1-1), \qquad Q\cdot P=Q^TLP=J\, ,
\ee
$\wt\Phi(\wt \rho,\wt \sigma, \wt v)$ is a known function of
three complex variables $(\wrh,\ws,\wv)$ and $C_>$ and $C_<$ are
a pair of three real dimensional subspaces of the
three complex dimensional space labelled by $(\wt\rho,\ws,\wv)
\equiv (\wrh_1+i\wrh_2,\ws_1+i\ws_2,\wv_1+i\wv_2)$.
They are defined as
\bea{ep2kkcomb}
C_>&:& 
\wt \rho_2=M_1, \quad \wt\sigma_2 = M_2, \quad
\wt v_2 = -M_3, \nonumber \\
&&  0\le \wt\rho_1\le 1, \quad
0\le \wt\sigma_1\le N, \quad 0\le \wt v_1\le 1\, ,
\nonumber \\
C_<&:&
\wt \rho_2=M_1, \quad \wt\sigma_2 = M_2, \quad
\wt v_2 = M_3, \nonumber \\
 && 0\le \wt\rho_1\le 1, \quad
0\le \wt\sigma_1\le N, \quad 0\le \wt v_1\le 1\, ,
\een
$M_1$, $M_2$ and $M_3$ being large but fixed positive
numbers with $M_3<< M_1, M_2$.  
For $\wt v\simeq 0$, 
$\wt\Phi$ takes the form:
\be \label{es1}
\wt\Phi(\wrh,\ws,\wv) = -4\pi^2\, \wt v^2 \, f(N\wrh) g(\ws / N)
+ \OO(\wt v^4)\, ,
\ee
where $(f(\tau))^{-1}$ and 
$(g(\tau))^{-1}$ have the interpretation of the generating
function for the degeneracies of
purely electric half-BPS states and purely magnetic
half-BPS states respectively. 
For example for the $\ZZZ_N$ orbifold of the heterotic string theory
on $T^4\times\wh S^1\times S^1$ with prime values of $N$ 
we have\cite{0510147}
\be \label{efgtau}
f(\tau) = (\eta(\tau/N))^{k+2} \eta(\tau)^{k+2}, \qquad
g(\tau) = (\eta(\tau))^{k+2} \eta(N\tau)^{k+2}\, , \qquad
k \equiv {24\over N+1}-2\, .
\ee
For $N=1$, \i.e.\ for heterotic string theory on 
$T^4\times \wh S^1\times S^1$, this gives us back the
standard result $\eta(\tau)^{24}$ for both $f(\tau)$ 
and $g(\tau)$.

The jump in the degeneracy as we move from 
$a_\infty<a_c$ to $a_\infty>a_c$ is determined by an integral over the
difference between the contours $C_>$ and $C_<$. 
The contribution to
this integral comes from the pole of the integrand
at $\wt v=0$\cite{0702141}. 
Substituting \refb{es1} into \refb{e8} and
evaluating the residue at the pole at $\wt v=0$ we get
\be \label{es2}
d_>(Q,P)- d_<(Q,P)
= - Q\cdot P \, d_{el}(Q) \, d_{mag}(P)\, ,
\ee
where
\be \label{es3}
d_{el}(Q) = \int_0^1d\wrh \, e^{-i\pi N\wrh Q^2} \left(f(N\wrh)
\right)^{-1}\, ,
\qquad
d_{mag}(P) ={1\over N}\int_0^N \, d\ws\,
e^{-i\pi \ws P^2 / N} 
\left( g(\ws/N)\right)^{-1}\, ,
\ee
are the degeneracies of purely 
electric and purely magnetic half-BPS states carrying
charges $Q$ and $P$ respectively. 
Thus $\ln d_{el}(Q)$ and $\ln d_{mag}(P)$ 
are the entropies of small
black holes of electric charge $Q$ and magnetic charge $P$
respectively. Since $\ln|Q\cdot P|$ is subleading compared to these
entropies for large $Q^2$ and $P^2$ 
\i.e.\ for
\be \label{enewrange}
n, Q_1, Q_5 >> 1\, ,
\ee
we see that $\ln |d_>(Q,P)- d_<(Q,P)|$ 
can be identified as the sum of the entropies of a small electric black
hole of charge $Q$ and a small magnetic black hole of charge $P$.
In carrying out the analysis on the black hole side we shall choose
charge vectors satisfying \refb{enewrange}.

Taking into account the sign of the right hand side of \refb{es2},
and assuming that this phenomenon has a description in the
dual black hole picture,
we can draw the following conclusion:\footnote{There are two
points to note here. First when a new configuration with same
charge appears in the black hole system, its degeneracy (or
more precisely the index), \i.e.\
exponential of the entropy, will add to the degeneracy of the
other configurations of the same charge. Second, we shall be
implicitly assuming that the new system that appears gives a
positive contribution to $d(\vec Q,\vec P)$. Otherwise the condition on
$Q\cdot P$ stated in the proposal will be reversed. With the
sign convention for $d(\vec Q,\vec P)$ described in footnote \ref{f1}
this assumption is consistent with the wall crossing formula of
\cite{0206072,0702146}.}

\noindent {\it For $J(=Q\cdot P)>0$, as we cross the wall of 
marginal stability
\refb{e6} from $a_\infty>a_c$ to $a_\infty<a_c$, 
a new configuration should
appear whose entropy is equal to the sum of the entropies of
a small electric black
hole of charge $Q$ and a small magnetic black hole of charge $P$.
On the other hand for $J(=Q\cdot P)<0$, 
as we cross the wall of 
marginal stability
\refb{e6} from $a_\infty<a_c$ to $a_\infty>a_c$, 
a new configuration should
appear whose entropy is equal to the sum of the entropies of
a small electric black
hole of charge $Q$ and a small magnetic black hole of charge $P$.}

\noindent In \S\ref{s3} we shall verify this explicitly by identifying the
new configuration as a two centered black hole solution
with an electric
center of charge vector $Q$ and a magnetic center of charge vector $P$.

\sectiono{Two Centered Small Black Holes} \label{s3}

For describing the two centered black hole we shall use the
$\NN=2$ supersymmetric description of the same system described
above. In the supergravity approximation the relevant part of the
theory is described by the prepotential (see \cite{0007195}
for a review):
\be \label{e2.1}
F = - {X^1 X^2 X^3\over X^0}\, ,
\ee
where $X^I$'s are scalar fields. These are related to the scalar
fields $a+iS$ and $M$ 
via the relations
\be \label{e2.2}
a+iS = {X^1\over X^0}, \quad T = -i{X^2\over X^0}, \quad 
U = -i{X^3\over X^0}\, ,
\ee
$iT$ and $iU$ being the Kahler and complex structure modulus
of the torus $\wh S^1\times S^1$. They contain the same information
as the matrix $M$. In particular for the asymptotic $M$ given in
\refb{e4}, we have
\be \label{e2.3}
T_\infty = R \wh R, \quad U_\infty = \wh R/R \, .
\ee
The theory contains four gauge fields, and we shall denote the 
electric and magnetic charges associated with these
gauge fields by 
$q_0,q_1,q_2,q_3$ and $p^0,p^1,p^2,p^3$ respectively.
These charges can be related to the charge vectors $Q$ and $P$
introduced earlier via the relation: 
\be \label{e2.4}
Q = \pmatrix{q_0\cr q_3\cr -p^1\cr q_2}, \qquad 
P = \pmatrix{q_1\cr p^2\cr p^0\cr p^3}\, .
\ee
Thus for the configuration \refb{e5} we have
\be \label{e2.5}
(q_0,q_1,q_2,q_3) = (0, Q_1-1, -1, -n/N),  \qquad
(p^0, p^1, p^2, p^3) = (Q_5, 0,-J,  0)\, .
\ee
The theory has an underlying gauge invariance that allows for a
scaling of all the $X^I$'s by a complex function. We shall fix this gauge
using the gauge condition:
\be \label{e2.6}
i (\bar X^I F_I - X^I \bar F_I) = {1}\, , \qquad F_I \equiv \p F
/ \p X^I\, ,
\ee
which amounts to setting $\alpha'=8$.
This fixes the normalization but not the overall phase 
of the $X^I$'s.
While studying a black hole solution carrying a given set
of charges, it will be convenient to fix the overall phase of
the $X^I$'s such that
\be \label{e2.7}
Arg(q_I X^I - p^I F_I) = \pi \quad \hbox{at $\vec r=\infty$}\, .
\ee
In this gauge one can construct 
a general multi-centered black hole solution with
charges $(q^{(s)}, p^{(s)})$ located at $\vec r_s$. 
The locations $\vec r_s$ are constrained by the 
equations\cite{0005049,0101135,0304094}
\be \label{e2.8a}
h_I p^{(s)I} - h^I q^{(s)}_I + \sum_{t\ne s} {
p^{(s)I} q^{(t)}_I - q^{(s)}_I p^{(t) I} \over |\vec r_s - \vec r_t|} = 0
\ee
where $h^I$ and $h_I$ are constants defined through the
equations
\be \label{e2.8}
X^I_\infty -\bar X^I_\infty = i  h^I \, , \qquad
F_{I\infty} - \bar F_{I\infty} = i   h_I \, .
\ee
If we
define $\alpha$ and $\beta$ via the relations
\be \label{e2.9}
X^0_\infty = \alpha+i\beta\, ,
\ee
then using \refb{e2.1}-\refb{e2.3} and \refb{e2.8} we get
\ben \label{e2.10}
&& h^0 = 2\beta, \quad h^1 = 2 (\beta a_\infty + \alpha S_\infty), \quad
h^2 = 2\wh R R \alpha, \quad h^3 = 2\wh R \alpha / R, 
\nonumber \\
&&
h_0 = -2\wh R^2 (\alpha S_\infty +\beta a_\infty),  
\quad h_1 = 2\beta \wh R^2,
\quad h_2 = 2 \wh R (\beta S_\infty -\alpha a_\infty) / R, 
\nonumber \\ &&
h_3 = 2\wh R R (\beta S_\infty - \alpha a_\infty)\, . 
\een
The gauge condition \refb{e2.6} gives
\be \label{e2.11}
\alpha^2 + \beta^2 = (8 \wh R^2 S_\infty)^{-1}\, .
\ee

To proceed further we need to focus on a specific multi-centered solution.
Since our goal is to identify a configuration whose entropy is the sum
of the entropies of a purely electric small black hole of charge $Q$ and
a purely magnetic small black hole of charge $P$, the natural object to
focus on is a two centered solution with electric charge $Q$ at one center
and a magnetic charge $P$ at the other center. This will automatically have
the desired entropy.\footnote{In the supergravity approximation the
solution is singular at each center, 
but once higher derivative corrections are
taken into account each center is transformed into the near horizon
geometry of a 
non-singular extremal black
hole with finite entropy equal to the statistical entropy of the
corresponding microstates. 
This has been demonstrated explicitly for the $\ZZZ_N$ orbifolds
of heterotic string theory on 
$T^4\times\wh S^1\times S^1$\cite{0409148,0410076,
0411255,0411272,0501014,0502126,0502157,
0507014,0506176}. In this case 
the modifications of the solution
due to higher derivative
corrections can be found using the method developed in
\cite{0009234}. This approach fails for type II string compactification,
most likely due to the absence of an $AdS_3$ factor in the near
horizon geometry of the small black hole. However it is expected that
once the effect of full set of higher derivative terms are taken into
account the entropy of a small black hole in type II string theory
will also reproduce the statistical entropy of the corresponding
microstates.}
Using \refb{e5}, \refb{e2.4} we see that the charges at the two centers
are given by:
\be \label{e2.12}
q^{(1)} = (0,0,-1,-n/N), \quad p^{(1)} = (0,0,0,0),
\quad q^{(2)} =(0, Q_1-1, 0,0), \quad p^{(2)} =(Q_5, 0, -J, 0)\, .
\ee
Eqs.\refb{e2.8a} for $s=1$ and 2 now gives:
\be \label{e2.13a}
h^2 + {n\over N} h^3 = {J\over L}, 
\ee
\be \label{e2.13b}
h_0 Q_5 - h_2 J - h^1 (Q_1-1)
+ {J\over L} = 0\, ,
\ee
where $L=|\vec r_1 -\vec r_2|$ is the separation between the two
centers.
Using \refb{e2.10} and 
\refb{e2.13a} we get
\be \label{e2.14}
\alpha = {J\over 2L} {1\over R\wh R + {n\over N}{\wh R\over R}}\, .
\ee
Using \refb{e2.10} and \refb{e2.14} 
we may now express
\refb{e2.13b} as
\be \label{e2.15}
\beta \left(a_\infty (Q_1 - 1 + \wh R^2 Q_5) + {\wh R J 
S_\infty \over R} 
\right) 
+ \alpha \left( (Q_1 - 1 + \wh R^2 Q_5) S_\infty -\wh R R -{n\over N}
{\wh R\over R} - {\wh R J a_\infty \over R}
\right) = 0
\, .
\ee
Substituting \refb{e2.14}, \refb{e2.15} into \refb{e2.11} we can
determine $L$. The ambiguity in determining the sign of $L$
can be fixed using \refb{e2.7}.

We are interested in determining under what conditions the two
centered black hole solution described above exists. For this we note
that a sensible solution should have positive value of $L$. Typically
as we change the values of the asymptotic moduli keeping the
charges fixed, the value of $L$ changes. On some subspace of 
codimension
1 the value of $L$ becomes infinite and beyond that the solution gives
negative values of $L$ which means that the solution does not exist.
To determine this codimension 1 subspace we simply need to determine
the conditions on the asymptotic moduli for which $L=\infty$. 
{}From \refb{e2.14} we see that in this case $\alpha=0$. 
Since eq.\refb{e2.11} now requires $\beta$ to be non-zero, we see from
\refb{e2.15} that
\be \label{e2.16}
a_\infty (Q_1 - 1 + \wh R^2 Q_5) + \wh R J S_\infty / R =0\, .
\ee
This is identical to the condition \refb{e6} for marginal 
stability\cite{0005049}. 
Thus we conclude that as $a_\infty$ 
passes through
$a_c$, the two centered black hole solution carrying an entropy
equal to the sum of the entropies of a small electric black hole of
charge $Q$ and a small magnetic black hole of charge $P$, (dis)appears
from the spectrum.
This is precisely what was predicted at the end of \S\ref{s2}
by analyzing the exact
formula for the degeneracy of dyons.

In order to complete the verification of the predictions made at the end
of \S\ref{s2} we need to determine on which side of the 
$a_\infty=a_c$ line
the two centered solution exists. 
For this we use eq.\refb{e2.7}. For the solution under consideration
this gives, using \refb{e2.15},
\be \label{esign}
\alpha \left(a_\infty (Q_1 - 1 + \wh R^2 Q_5) + {\wh R J S_\infty 
\over R} 
\right) \left\{ 1 + 
{\left( (Q_1 - 1 + \wh R^2 Q_5) S_\infty -\wh R R -{n\over N}
{\wh R\over R} - {\wh R J a_\infty \over R}
\right)^2 \over  \left(a_\infty (Q_1 - 1 + \wh R^2 Q_5) + {\wh R J S_\infty 
\over R} 
\right)^2} \right\} < 0\, .
\ee
First consider the case $J>0$. 
Since $L$ must be positive for the two centered solution to
exist, we see from \refb{e2.14} that $\alpha>0$. 
In this case the term on the left hand side of \refb{esign} is
negative for $a_\infty<a_c$ and positive for 
$a_\infty>a_c$. Thus the inequality
is satisfied only for $a_\infty<a_c$, leading to the conclusion that the
two centered black hole exists only for $a_\infty<a_c$. A similar analysis
shows that for $J<0$, the two centered black hole exists only for
$a_\infty>a_c$. This is exactly what has been predicted at the end of
\S\ref{s2} from the analysis of the exact dyon spectrum of the theory.

\sectiono{Conclusion} \label{s4}

The main conclusion that can be drawn from the analysis of this
paper is that the exact formula for the degeneracy of dyons in
$\NN=4$ supersymmetric string theories encodes information
not only about the single centered black holes, but also about the
multi-centered black holes whose total charge adds up to that of
the dyon whose degeneracy is under consideration. Since in the
present example the contribution to the degeneracy from the
two centered black holes is exponentially small compared
to that from the 
single centered black hole, our results indicate that the
correspondence between the microscopic degeneracy of states and
black hole entropy extends beyond the leading asymptotic expansion,
-- not only for terms which are suppressed by inverse powers of
charges but also for terms which are exponentially suppressed.

\bigskip

Note added: The relation between the two centered black holes and the jump 
in the degeneracy in $\NN=4$ dyon spectrum has also been discussed in 
\cite{0706.2363} which appered a few days after this paper was first 
submitted to the arXiv.

\end{document}